# Understanding the role of rock heterogeneity in controlling fault strength and stability

Shaobo Han[1], Xiaoying Zhuang[1]*, Quanzhou Yao[1], Qianlong Zhou[2], Xiaodong Hu[2]*

[1]Department of Mathematics and Physics, Leibniz University Hannover, Hannover, 30167, Germany.
[2]Unconventional Petroleum Research Institute, China University of Petroleum-Beijing, Beijing, 102249, China.
*e-mail: zhuang@iop.uni-hannover.de

**Abstract**   The rock heterogeneity exists widely in fault zones; however, the intrinsic mechanism of how it affects the mechanical behavior of faults is poorly understood. To develop a quantitative understanding of the effect of the rock heterogeneity on the strength and stability of faults, here we investigate a pore-pressure model based on rate- and-state friction in the manner of two-degree-of-freedom spring-sliders and analyze the reasons of fault weakening and the conditions of frictional instability by carrying out nonlinear simulations and a linear stability analysis. We find that the strength of heterogeneous faults depends largely on the compaction difference (or differential compaction) between the two gouges (e.g. quartz and clay), and the stability is affected by the proportion of the two gouges patches. Our model implies that the rock heterogeneity is likely to weaken faults and reduce the stability of faults.

**Key Points**:

- A two-degree-of-freedom spring-sliders model based on rate- and-state friction considering the interaction of pore pressure between the two gouges is proposed to describe the effect of the rock heterogeneity on the strength and stability of faults.
- The differential compaction between the two gouges is the key mechanism of faults weakening.
- The rock heterogeneity can reduce the stability of faults, which is largely affected by the proportion of the two gouges patches.

**Plain Language Summary**   We have believed that heterogeneous gouges are common in fault zones, which generally consists of quartz and clay, and such heterogeneity can promote seismic slip propagation. Recent experiments show that gouge heterogeneity is hypothesized to be the cause of reduced fault strength and stability. To prove and explain this hypothesis in more detail, we propose a simple model based on the fundamental laws of friction physics. We find that the difference in thickness between different fault gouges reduces the strength of the fault, but unlike the experiment, this difference in thickness is fluctuating, which may cause the fluctuating change in the strength of the fault. In addition, the heterogeneity of fault gouges increases the risk of fault instability. These phenomena suggest that gouge heterogeneity may be a key mechanism to reduce fault strength and stability.

## 1. Introduction

Geological heterogeneity is abundant in complex crustal fault zones (Tesei et al., 2014; Ando et al., 2010; Wei et al., 2013; Yamashita et al., 2021; Saffer et al., 2015; Tal et al., 2018), this heterogeneity is inherent to faults and represents different fault structures, rheology, and rock combinations (such as mixed gouges composed of quartz, clay, etc.). (Hawthorne & Rubin, 2013; Saffer & Wallace, 2015; Ando et al, 2023; Kirkpatrick et al., 2021). The heterogeneity of gouges with different physical and frictional properties, as one of the heterogeneity, may lead to changes in the fault strength (Thomas et al., 2018; Skarbek et al., 2012; Ando et al, 2023; Feng et al, 2023) and fault stability (Bedford et al., 2022). The low dynamic frictional resistance along the fault promotes the slip of the fault, which is generally explained by phyllosilicates (such as clay) weakening the fault strength and affecting the fault stability (Bedford et al., 2022; Feng et al, 2023). Some experimental and numerical results show the additional clay significantly weakens the shear strength of quartz-rich fault gouge (Smeraglia et al., 2017; Wang et al., 2017). In addition, a series of slip behaviors carried by faults, such as seismic slip and aseismic creep (Avouac, 2015; Thomas et al., 2018; Skarbek et al., 2012), slow slip and slow seismic (Barnes et al., 2020; Kirkpatrick et al., 2021; Bürgmann, 2018), are generally considered to be caused by the heterogeneity of mixed rocks in fault zones (Collettini et al., 2019; Barnes et al., 2020; Bedford et al., 2022). Although attempts have been made to explain the observations using discrete element models (Wang et al., 2017; Wang et al., 2019) and highly simplified elastic mechanics model (Bedford et al., 2022), the underlying physical mechanism of how gouge heterogeneity affects fault slip characteristics is still poorly understood.

Multi-degree-of-freedom spring-slider models based on rate-state law have been used to study the effect of geological heterogeneity on fault stability and slip behavior. These models simplify asperity and other heterogeneity into sliders with rate-strengthening and rate-weakening (Yoshida & Kato, 2003; Skarbek et al., 2012; Luo & Ampuero, 2017; Luo & Liu, 2021; Ando et al, 2023); however, the models do not account for differences in compaction/ dilatancy evolution between different rocks,and the differences of normal stress distribution in each patch. Considering the differential compaction between fault gouges during shear and the resulting effective normal stress redistribution is important for describing fault strength evolution and fault stability, as demonstrated by the different evolution of layer thickness and volume between quartz and clay (Bedford et al., 2022). Here we develop a pore-pressure model of earthquake nucleation in the manner of two-degree-of-freedom spring-sliders based on rate- and-state friction law, which takes into account the differential thickness evolution between fault gouges, and the interaction of pore pressure in different patches. Finally we investigate the effects of fault rock heterogeneity on the fault strength and stability.

Our model shows that the strength of heterogeneous faults depends largely on the differential compaction between the two gouges. This vertical compaction difference results in the redistribution of the effective normal stress acting on the two gouges, and finally the friction strength of the fault shows a gradual weakening. In addition, the rock heterogeneity reduces the stability of faults. Our finding is qualitatively consistent with laboratory observations of Bedford et al., 2022.

## 2. Results and Discussion

### 2.1. Two-degree-of-freedom pore-pressure spring-slider model

When the fault slips, because of the compaction difference between different gouges, the normal stress acting on the fault redistributes, and finally the strength and stability of faults change. To model the influence of the rock heterogeneity on the strength and stability of the fault, we develop a two-degree-of-freedom pore-pressure spring-slider model (Fig. 1).

Our model consists of two sliders that are pulled by two springs that are restricted to moving at a steady slip rate. The bottom of the sliders is the granular pack, representing different fault gouges, which are deformable. Different gouges have different physical properties, and this maybe lead different compaction evolution between gouges. The base area of the two sliders is the same or different, which approximately represents the content or fraction of the two gouges. The spring between the sliders represents the interconnection between fault areas. The damage zone represents a permeability or drained boundary with a specific diffusion length. The model considers the interaction of pore pressure between the two gouges.

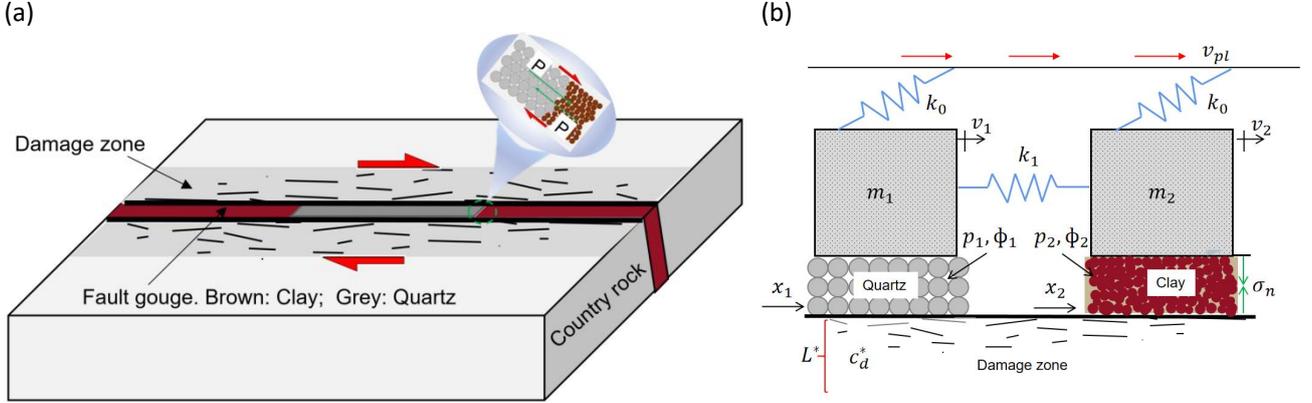

**Figure 1.** Conceptual picture of fault rock heterogeneity including pore-pressure effects. (a) Schematic diagram of a typical heterogeneous fault zone. The green arrow represents the diffusion of pore pressure (P) between the two gouges. (b) Our two-degree-of-freedom spring-slider model. $m_1$ and $m_2$ are the mass of block-1 and block-2 respectively. Block-1 and block-2 are connected by spring $k_1$, which is a coupling spring, an interconnection between fault regions. $k_0$ is the connection between the plate tectonic loading with a constant drive at the far end and the fault. Plate tectonic processes move at a constant rate $V_{pl}$, $V_1$ and $V_2$ are the velocity of each slider, $p_1$ and $p_2$ are the pore pressure in the gouge of block-1 and block-2 respectively, $c_d^*$ is the hydraulic diffusion coefficient in the damage zone, L* is characteristic diffusion length, $\phi_1$ and $\phi_2$ are the porosity in the gouge of block-1 and block-2 respectively.

According to Newton's second law, the motion equations of the spring-slider system in the horizontal direction are obtained (Appendices A1.1). The frictional shear stress is expressed by the equation $\tau_i = \mu(V_i, \Theta_i)(\sigma - p_i)$, where the friction coefficient μ on the fault surface depends on the slip rate V and state variable Θ, i=1, 2. The state evolution adopts the "aging law" version (Segall and Rice, 1995)(the other is the slip type), and takes into account the change of effective normal stress (Appendices A1.2). Here we introduce the double spring-slider model to describe the two gouges. Different from previous two-degree-of-freedom models, we consider dilatation/compaction and pore pressure changes within each patch, as well as the interaction of pore pressure between different fault gouges (Eq. (1-4)). (Note: Only pore pressure and porosity evolution equations are shown here, See appendices information for the rest of the equations):

$$\frac{dp_1}{dt} = c_{d1}^*(p^\infty - p_1) - \frac{\dot{\phi}_1}{\beta_1} + \frac{A_2}{A_1+A_2}\frac{dp_2}{dt} \tag{1}$$

$$\frac{dp_2}{dt} = c_{d2}^*(p^\infty - p_2) - \frac{\dot{\phi}_2}{\beta_2} \tag{2}$$

$$\frac{d\phi_1}{dt} = -\frac{V_1}{L_1}(\phi_1 - \phi_{ss1}) \tag{3}$$

$$\frac{d\phi_2}{dt} = -\frac{V_2}{L_2}(\phi_2 - \phi_{ss2}) \tag{4}$$

where $c_{di}^*$ ( $= \kappa/\nu\beta d^2$) is the hydraulic diffusion coefficient in the damage zone, $\kappa$ is the permeability, $\nu$ is the fluid viscosity, d is characteristic diffusion length, $p^\infty$ is the initial pore pressure or ambient pressure, $p_i$ is the pore pressure in the gouge, $\beta_i$ is coefficient of compressibility, $A_i$ represents the content or fraction of the gouges, $L_i$ is the characteristic slip distance, $V_i$ is the velocity of each slider, $\phi_i$ is the porosity in the gouge, $\phi_{ssi}$ is the porosity in the gouge at steady state (i=1, 2).

## 2.2. The influence of rock heterogeneity on fault strength

The fault strength is strongly dependent on the rock heterogeneity (Smeraglia et al., 2017; Wang et al., 2017). For a heterogeneous fault, the combination of quartz (rate-weakening material) and clay (rate-strengthening material) results in a higher likelihood of the weakening of fault strength.

To understand how rock heterogeneity may influence the strength of fault, we simulate the shearing under different clay fractions (See Supplementary Note 1 for initial conditions and parameters), and examine the frictional coefficient of each. Fig. 2 demonstrates how clay fraction influences the evolution of the fault strength. We observe that heterogeneous faults show a universal weakening trend, with friction strength evolving towards the value of $\mu_{pure\ clay}$ (pure clay, μ=0.3), which is qualitatively consistent with the experimental results of Bedford (2022). Meantime, we calculate the total friction strength $\mu_{total}(= f_{clay}\mu_{clay} + (1 - f_{clay})\mu_{quartz})$ of different clay fractions respectively. The results show that the frictional coefficient of the system decreases as a whole, which indicates the frictional resistance of heterogeneous faults continues to reduce with the increasing of the clay fraction (Fig. 3). To ensure that this weakening is caused by the heterogeneity itself and not by the arrangement of patches, we also perform symmetry inversion of quartz and clay patches. The fault before and after inversion shows similar weakening in the tests (Fig. 4), suggesting the fault weakening is indeed induced by rock heterogeneity itself.

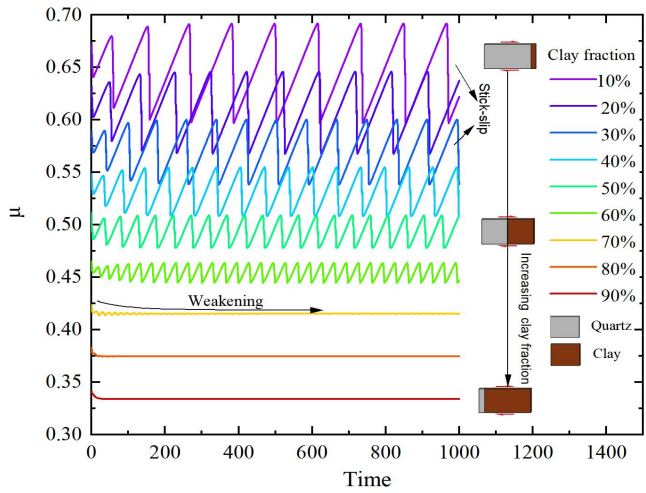 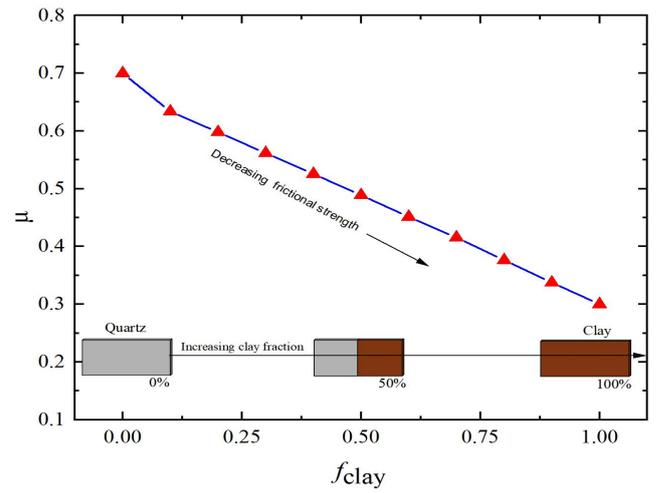

**Figure 2.** The evolution of the frictional coefficient with time for the heterogeneous faults.

**Figure 3.** The mean frictional coefficient under different clay fractions.

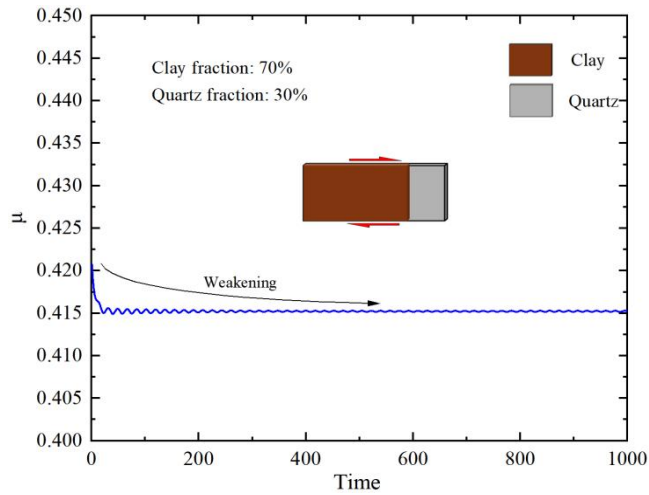

**Figure 4.** The evolution of the friction coefficient (μ) with time (Clay, 70%; Quartz, 30%. Symmetry reversal of heterogeneous gouge)

Here we give the pore volume strain and effective normal stress evolution data from quartz and clay gouges (20% quartz + 80% clay), which shows that the compaction difference between the two gouges rapidly increases to about 0.64% of the layer thickness (H). At the same time, the effective normal stress difference ($\sigma_{clay}^{eff} - \sigma_{quartz}^{eff}$) on clay and quartz increases rapidly, then decreases and fluctuates, and the difference tends to be zero, which indicates that the differential compaction results in the redistribution of the effective normal stress acting on the two gouges (Fig. 5). We think that the friction resistance of faults may become weaker because the weaker clay gouge supports greater effective normal stress. Based on above, we calculate this friction coefficient using a simplified elastomechical method (Knuth et al., 2013; Bedford et al., 2022) whose calculation depends on the elastic modulus. The additional strain $\Delta\varepsilon$ of the gouge will correspond to the additional effective normal stress $\Delta\sigma = \left(K_{bulk} + \frac{4G_{shear}}{3}\right)\Delta\varepsilon$ acting on the gouge (Knuth et al.,

2013) , $K_{bulk}$ is the bulk modulus, and $G_{shear}$ is the shear modulus. The results show that differential compaction leads to the potential weakening effecting dependent on the elastic modulus of the gouge, but they are constrained better only in the late stage (Fig. 6, the red and blue dotted curve). However, this constrain gradually deteriorates with the increasing of quartz fraction, but it is still within the range of theoretical friction strength evolution (See Supplementary Figure 1 for more constraints with different clay fractions), which shows that the effective normal stress redistribution due to differential compaction can explain the weakening of most of rock heterogeneity faults. Therefore, the results suggest that the differential compaction between the two gouges, and the associated effective normal stress redistribution, may weaken the fault.

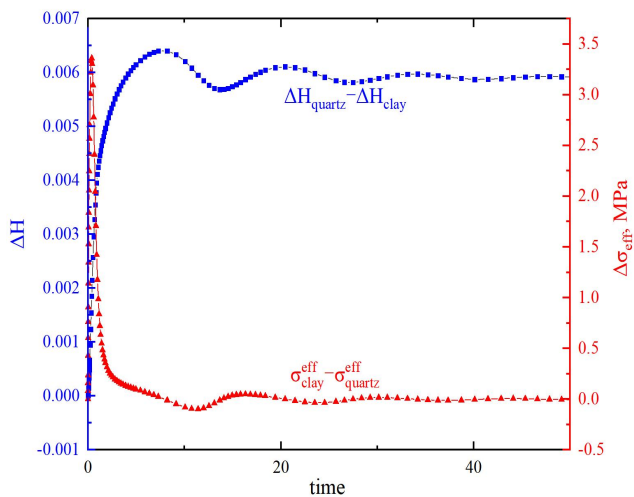
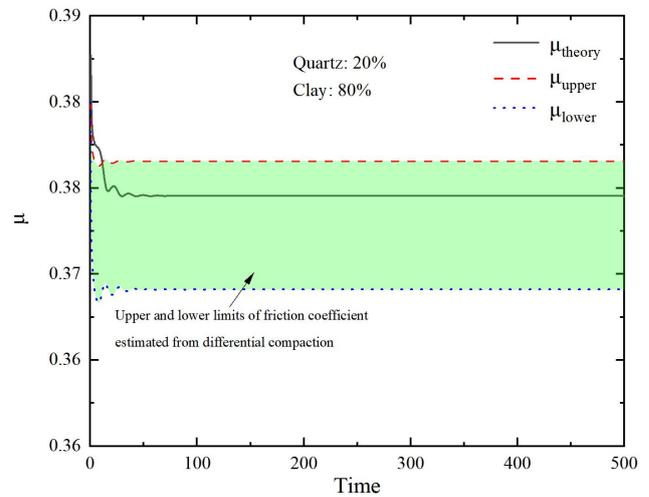

**Figure 5.** The compaction difference and effective normal stress difference between the quartz and clay gouges, $\Delta H = \Delta \phi * H_{layer}$.

**Figure 6.** The potential weakening from the differential compaction between the quartz and clay gouges.

## 2.3. Stability analysis

When an infinitesimal perturbations is applied to the steady-state velocity, the stability of the stable frictional sliding determines whether the motion is a slow steady sliding or a violent stick-slip (Skarbek et al., 2012; Luo & Ampuero, 2017; Alghannam & Juanes, 2020). Generally, at a homogeneous fault and a constant pore pressure, linear stability analysis of the system about steady-state leads to the stability condition by Segall (1995). However, for heterogeneous faults, the pore pressure evolves with differential compaction between the two gouges. To gain an understanding of the stability of rock heterogeneity faults systems, we investigate the linear stability of a simplified two-degree-of-freedom spring-slider system (See appendices A2). Because the analytical solution of the linear stability equation is quite complicated, we only give the numerical solution. Linear stability analysis shows that the heterogeneous fault will slip unsteadily when the shear stiffness of the loading system is lower than a minimum critical value $(k < k_{0crit}^{min})$, and the fault will slip stably when the shear stiffness is greater than a maximum critical value $(k > k_{0crit}^{max})$. When the shear stiffness is between the maximum and minimum critical values $(k_{0crit}^{min} < k < k_{0crit}^{max})$, the system is in a metastable state, in which the system is between stable and unstable. At this time, if the factors that determine the state of the system change, the system will develop in a stable or unstable direction.

As the fraction of clay moves away from the two end-elements 0 and 100%, the instability zone gradually increases, which indicates that the rock heterogeneity increases the probability of fault instability (producing an overall reduction in the stability of faults) (Fig. 7). It should be noted that the sliding instability of rock heterogeneity faults is largely affected by the proportion of different gouges, especially adequate amounts of quartz. The unstable areas remained at a high level when the fraction of clay (weak rate-strengthening patch) is less than 30% or the fraction of quartz (strong rate-weakening patch) is greater than 70%, and the stick-slip instability is likely to occur obviously (Fig. 2).

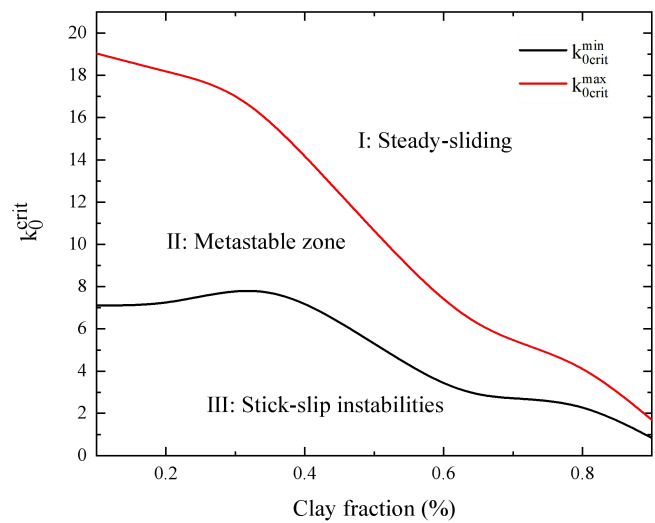

**Figure 7.** The critical stiffness of the system under different clay fractions

## 2.4. Compaction evolution under variable effective normal stress

The variation of the effective normal stress affects the compaction evolution of gouges (Linker & Dieterich, 1992; Segall & Rice, 1995). Different from the homogeneous fault, differential compaction in the heterogeneous fault results in the redistribution of the effective normal stress, which makes the compaction of the two gouges produce different evolution. Here we model the shearing friction along the heterogeneous fault under variable effective normal stress, and capture the compaction/dilatation effect of gouges, which differs from the compaction evolution of gouges under constant effective normal stress experimentally (Bedford et al., 2022).

The layer thickness evolution of the two gouges layers is back-calculated by using the porosity evolution data. We assume that the sliding area is constant, and the differential vertical strain between the two gouges is caused by the differential inelastic compression applied to the two gouges. This vertical differential strain results in a change in the effective normal stress distribution acting on the two gouges. We observe that the quartz gouge experiences more shear compaction than the clay gouge (Fig. 8), which means that the weaker clay supports a greater effective normal stress during compaction (Fig. 9). According to the estimation (Fig. 6) and simulation (Fig. 2) of the friction coefficient, the friction strength of the fault shows a gradual weakening. However, the compaction difference is different from the results observed in experiments (Bedford et al., 2022), but shows a periodic fluctuation related to compaction/dilatancy. This change results in a periodic redistribution of the effective normal stress acting on the two gouges.

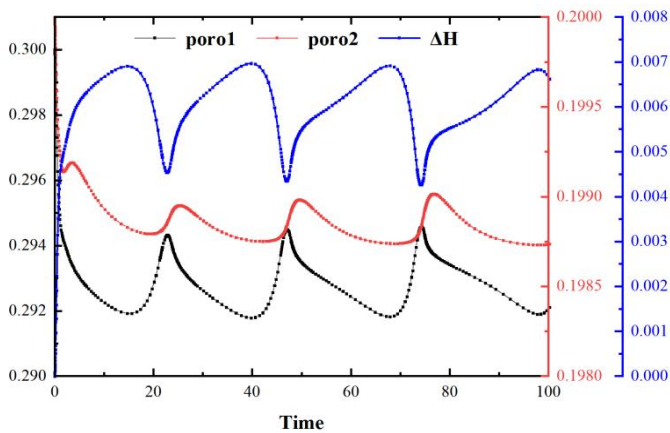 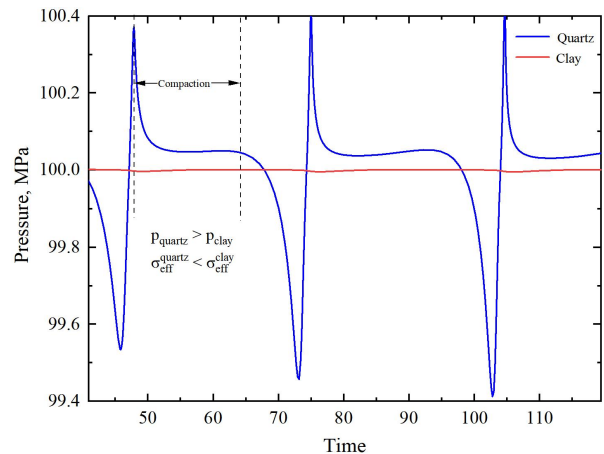

**Figure 8.** The evolution of the pore volume and the difference of thickness reduction of the quartz (poro1) and clay (poro2). ΔH is the difference of thickness reduction(=ΔH$_{quartz}$ - ΔH$_{clay}$). (40% quartz +60% clay)

**Figure 9.** The pore pressure acting on the two gouges. (40% quartz +60% clay)

## 3. Conclusion

We develop a pore-pressure model based on rate- and-state friction in the manner of two-degree-of-freedom spring-sliders. The model captures and characterizes the compaction/dilation effects of gouges, and takes into account the interaction of pore pressure between different gouges under variable normal stress. The results of the strength and stability of the heterogeneous fault during the shearing are simulated and analyzed. Our model suggests that differential compaction between different gouges is a key mechanism of fault weakening, and that rock heterogeneity leads to an overall reduction in fault stability.

## Conflict of Interest

The authors declare no competing interests.

# Appendices

## A1 Derivation of governing equations

### A1.1 Motion constitutive equation.

We assume that the basement area of block-1 is $A_1$ and the basement area of block-2 is $A_2$. $M_1$ and $M_2$ are the mass of block-1 and block-2 respectively. Block-1 and block-2 are connected by spring $K_1$, which is a coupling spring, an interconnection between fault regions, and an axial spring of strike-slip fault. $K_0$ is the connection between the plate tectonic loading with a constant drive at the far end and the fault. Plate tectonic processes move at a constant rate $V_{pl}$. $x_1$ and $x_2$ are the position coordinates of each slider, and $\tau_1$ and $\tau_2$ are the shear stress of each slider. Therefore, based on the second Newton's law, we get the motion equations of the sliders in the horizontal direction:

$$M_1 \ddot{x}_1 = K_0(V_{pl}t - x_1) + K_1(x_2 - x_1) - \tau_1 A_1 \tag{A.1}$$

$$M_2 \ddot{x}_2 = K_0(V_{pl}t - x_2) - K_1(x_2 - x_1) - \tau_2 A_2 \tag{A.2}$$

Here, we let

$$U_1 = V_{pl}t - x_1 \tag{A.3}$$

$$U_2 = V_{pl}t - x_2 \tag{A.4}$$

$$U_3 = x_2 - x_1 \tag{A.5}$$

where $U_1$, $U_2$ and $U_3$ are the relative displacement between the load point and slider 1, the load point and slider 2, and the slider 1 and slider 2, respectively. Taking the derivative of both sides at the same time, we obtain the differential equations for the relative displacement:

$$\dot{U}_1 = V_{pl} - V_1 \tag{A.6}$$

$$\dot{U}_2 = V_{pl} - V_2 \tag{A.7}$$

$$\dot{U}_3 = V_2 - V_1 \tag{A.8}$$

where $v_1$ and $v_2$ are slip velocity of slider-1 and slider-2 respectively. Rewrite the above equations of the motion (A.1) and (A.2) as:

$$\dot{V}_1 = \frac{1}{(T/2\pi)^2}\left[U_1 + \frac{k_{11}}{k_{01}}U_3 - \frac{1}{k_{01}}\left(\mu_1^* + \hat{a}_1 ln\frac{V_1}{V^*} + \hat{b}_1 ln\frac{\theta_1}{\theta^*}\right)(\sigma_n - p_1)\right] \tag{A.9}$$

$$\dot{V}_2 = \frac{1}{(T/2\pi)^2}\left[U_2 - \frac{k_{12}}{k_{02}}U_3 - \frac{1}{k_{02}}\left(\mu_2^* + \hat{a}_2 ln\frac{V_2}{V^*} + \hat{b}_1 ln\frac{\theta_2}{\theta^*}\right)(\sigma_n - p_2)\right] \tag{A.10}$$

where the shear stress along the base of each block is $\tau_i = \mu_i \sigma_n^{eff} = \left(\mu_i^* + \hat{a}_i ln\frac{V_i}{V^*} + \hat{b}_i ln\frac{\theta_i}{\theta^*}\right)(\sigma_n - p_i)$, the mass per unit area is $m_i = (T/2\pi)^2 k_0/A_i$, $T$ is the vibration period of an analogous free-sliding spring-slider system (Rice & Tse, 1986), $k_{0i} = K_0/A_i$, $k_{1i} = K_1/A_i$, $\hat{a}_i$ and $\hat{b}_i$ are the frictional parameters, i=1, 2.

### A1.2 State variable equations.

Here we use the "aging type" of the state evolution law (the other is slip type) in this paper, and take into account changes in effective normal stress (Linker and Dieterich, 1992; Segall and Rice, 1995):

$$\dot{\theta}_1 = 1 - \frac{\theta_1 V_1}{L_1} - \frac{\hat{\alpha}\theta_1(\dot{\sigma_n - p_1})}{\hat{b}_1(\sigma_n - p_1)} \tag{A.11}$$

$$\dot{\theta}_2 = 1 - \frac{\theta_2 V_2}{L_2} - \frac{\hat{\alpha}\theta_2(\dot{\sigma_n - p_2})}{\hat{b}_2(\sigma_n - p_2)} \tag{A.12}$$

where $\theta_i$ is the state variable, $V_i$ is the slip rate of the slider, $L_i$ is the characteristic slip distance, $\hat{\alpha}$ is a scaling parameter, p1 and p2 are the pore pressure in the gouge of slider-1 and slider-2 respectively, $\sigma_n$ is the normal stress. Simplify the above equations to:

$$\dot{\theta}_1 = 1 - \frac{\theta_1 V_1}{L_1} + \frac{\hat{\alpha}\theta_1 \dot{p}_1}{\hat{b}_1(\sigma_n - p_1)} \tag{A.13}$$

$$\dot{\theta}_2 = 1 - \frac{\theta_2 V_2}{L_2} + \frac{\hat{\alpha}\theta_2 \dot{p}_2}{\hat{b}_2(\sigma_n - p_2)} \tag{A.14}$$

### A1.3 Pore pressure constitutive equations.

The pore pressure equations adopt the equation proposed by Segall and Rice (1995),

$$\dot{p}_1 = c_{d1}^*(p^\infty - p_1) - \frac{\dot{\phi}_1}{\beta_1} \tag{A.15}$$

$$\dot{p}_2 = c_{d2}^*(p^\infty - p_2) - \frac{\dot{\phi}_2}{\beta_2} \tag{A.16}$$

where $c_{di}^* = \kappa/\nu\beta L^{*2}$ is the hydraulic diffusion coefficient in the damage zone, $\kappa$ is the permeability, $\nu$ is the fluid viscosity, $L^*$ is characteristic diffusion length, $p^\infty$ is the confining pressure, $\beta_i$ is coefficient of compressibility.

However, the two gouges have different pore pressure changes due to different compaction evolution, and there is no obvious physical barrier between the components, which makes the pore pressure in the two gouges interact with each other. Generally, the pressure difference causes the pore pressure in the clay to diffuse towards the quartz gouge, and the pore pressure in the quartz to diffuse towards the clay gouge otherwise. Therefore, we modify the pore pressure equation in quartz gouge, and the equation becomes:

$$\dot{p}_1 = c_{d1}^*(p^\infty - p_1) - \frac{\dot{\phi}_1}{\beta_1} + \frac{A_2}{A_1+A_2}\dot{p}_2 \tag{A.17}$$

where $\frac{A_2}{A_1+A_2}$ is approximately the mass fraction of clay or the proportion of clay in the total fault gouge. The equation links the pore pressure evolution of the two gouges.

**A1.4 Porosity constitutive equation.**

By analogy with the state evolution equation (Segall and Rice, 1995), a simple porosity evolution equation is considered:

$$\dot{\phi}_1 = -\frac{V_1}{L_1}(\phi_1 - \phi_{ss1}) \tag{A.18}$$

$$\dot{\phi}_2 = -\frac{V_2}{L_2}(\phi_2 - \phi_{ss2}) \tag{A.19}$$

where $\phi_{ssi}$ $(= \phi_0 + \epsilon ln\frac{V}{V_0})$ is the porosity at steady state, which only depends on velocity. $\epsilon$ is a "dilatancy coefficient." Because the porosity $\phi$ ranges from 0 to 1, here based on the general expression of the steady state porosity (Segall and Rice, 1995), and considering the porosity size within a certain range, we obtain the steady-state porosity of the two gouges:

$$\phi_{ssi} = \epsilon ln\left[\frac{d_{i1}V + d_{i2}}{d_{i3}+1}\right] \tag{A.20}$$

where $d_{11}$=5*exp(256.41); $d_{12}$=exp(85.47); $d_{13}$=5; $d_{21}$=5*exp(256.41); $d_{22}$=exp(85.47); $d_{23}$=5.

**A2 Linear stability analysis**

**A2.1 Linear stability analysis**

A common method for stability analysis with time-varying state variables is the quasi-steady-state approximation way (Lick, 1965; Rao & Arkin, 2003). Based on this approach, the time in the pore pressure solution is frozen, and then a linear stability analysis of the spring–slider system at a fixed pore pressure is performed (Alghannam & Juanes, 2020). The equations of motion at the quasi-steady-state are

$$\tau_1 A_1 = K_0(V_{pl}t - x_1) + K_1(x_2 - x_1) \tag{A.21}$$

$$\tau_2 A_2 = K_0(V_{pl}t - x_2) - K_1(x_2 - x_1) \tag{A.22}$$

The shear stress $\tau_i$ depends on the effective normal stress, and frictional coefficient, which is a function of the state variable $\theta_i$ and the slip velocity $V_i$, so

$$\tau_i = \sigma F(\theta_i, V_i) \tag{A.23}$$

In addition, the rate of change of the state variable also depends on the state variable $\theta_i$ and the slip velocity $V_i$, so

$$\dot{\theta}_i = G(\theta_i, V_i) \tag{A.24}$$

Since we only are interested in small perturbations near a steady state, we linearize so that the perturbations yield:

$$\tau_i^* = \sigma(F_{iv}V_i^* + F_{i\theta}\theta_i^*) \tag{A.25}$$

$$\dot{\theta}_i^* = G_{iv}V_i^* + G_{i\theta}\theta_i^* \tag{A.26}$$

where subscripts v and θ represent partial derivatives with respect to that variable. Small perturbations are marked with *. For example, $m_i \approx m_i^{ss} + q_i^*$.

We take the time derivative of Eq. (A.25), combining with Eq. (A.26), and finally rewrite the Eq. (A.25) in terms of velocity V:

$$\dot{\tau}_i^* = \sigma F_{iv}\dot{V}_i^* + \sigma(F_{i\theta}G_{iv} - F_{iV}G_{i\theta})V_i^* + G_{i\theta}\tau_i^* \tag{A.27}$$

We take the time derivative of Eq. (A.27), and finally get the following:

$$\ddot{\tau}_i^* = \sigma F_{iv}\ddot{V}_i^* + \sigma(F_{i\theta}G_{iv} - F_{iV}G_{i\theta})\dot{V}_i^* + G_{i\theta}\dot{\tau}_i^* \tag{A.28}$$

Substitute (A.21), (A.22) and their derivatives into (A.28). Finally, we get the system of linear equations:

$$0 = \sigma F_{1v}\ddot{V}_1^* + \left(\sigma F_{1\theta}G_{1v} - \sigma F_{1v}G_{1\theta} + \frac{K_0+K_1}{A_1}\right)\dot{V}_1^* - \frac{K_1}{A_1}\dot{V}_2^* + \frac{G_{1\theta}}{A_1}[-(K_0+K_1)V_1^* + K_1V_2^*] \tag{A.29}$$

$$0 = \sigma F_{2v}\ddot{V}_2^* + \left(\sigma F_{2\theta}G_{2v} - \sigma F_{2v}G_{2\theta} + \frac{K_0+K_1}{A_2}\right)\dot{V}_2^* - \frac{K_1}{A_2}\dot{V}_1^* + \frac{G_{2\theta}}{A_2}[-(K_0+K_1)V_2^* + K_1V_1^*] \tag{A.30}$$

The above equations can be written using matrix notation as:

$$M\ddot{\mathbf{V}} + N\dot{\mathbf{V}} + C\mathbf{V} = \mathbf{0} \tag{A.31}$$

We look for the solutions in the form $\mathbf{V} = \mathbf{V}_0 e^\lambda$, so:

$$Q(\lambda)\mathbf{V} = 0 \tag{A.32}$$

The stability of the linearized system depends on the signs of the real parts of the eigenvalues. We set det(Q)=0 and solve for λ:

$$det(Q) = r_4\lambda^4 + r_3\lambda^3 + r_2\lambda^2 + r_1\lambda + r_0 \tag{A.33}$$

We determine the stability boundary of the system by the sign change of the real part of λ. Therefore, we only need to look for the roots whose real part is 0. Finally, this will cause Eq. (A.33) to become (Luo & Ampuero, 2017):

$$r_1(r_1r_4 - r_2r_3) + r_0r_3^2 = 0 \tag{A.34}$$

Because the analytical solution of the Eq. (A.34) is very complicated, we only obtain the numerical solution.

# Supplementary Information

**Supplementary Note 1**

To understand the effect of rock heterogeneity on the fault strength and stability, we simulate the system equation of the spring-slider under the following initial conditions:

$$u_i(t=0) = (\mu_i^* + (\hat{a}_i - \hat{b}_i)ln(\frac{v_0}{v^*}))(\sigma_n - p_0)/k_{0i} \quad \text{(S1)}$$

$$v_i(t=0) = v_l \quad \text{(S2)}$$

$$\theta_i(t=0) = L/v_l \quad \text{(S3)}$$

$$p_i(t=0) = p_0 \quad \text{(S4)}$$

$$\phi_i(t=0) = \phi_0 \quad \text{(S5)}$$

where i=1, 2.

Here we solve the coupled two-degree-of-freedom pore-pressure spring-slider rigid system equations by nonlinear ordinary differential equation solver ode15s in Matlab. The simulation parameters are as follows: $a_1$=0.01; $b_1$=0.015; $a_2$=0.02; $b_2$=0.01; $\mu_1$=0.7; $\mu_2$=0.3; f=250 MPa; $p_0$=100 MPa; $v_l$=0.045 m/year; $\theta_1$=1.8; $\theta_2$=0.91; $L_1$=0.081 m; $L_2$=0.041 m; $\alpha$=0.2-1; $\beta_1$=10^(-4) 1/MPa; $\beta_2$=0.5*10^(-1) 1/MPa; $\epsilon_1$=0.00115; $\epsilon_2$=0.001; $c_1$=20; $c_2$=0.1. The values of $k_{0i}$ and $k_{1i}$ are in the Supplementary Table 1.

Supplementary Table 1. The value of $k_{0i}$ and $k_{1i}$ for different clay fractions
($K_0$=2.5×10$^6$ N/m, $K_1$=0.5×10$^6$ N/m)

| List | Quartz/% | Clay/% | $k_{01}$ | $k_{02}$ | $k_{11}$ | $k_{12}$ |
|---|---|---|---|---|---|---|
| 1 | 0.6 | 0.4 | 4.17 | 6.25 | 0.83 | 1.25 |
| 2 | 0.4 | 0.6 | 6.25 | 4.17 | 1.25 | 0.83 |
| 3 | 0.5 | 0.5 | 5.00 | 5.00 | 1.00 | 1.00 |
| 4 | 0.7 | 0.3 | 3.57 | 8.33 | 0.71 | 1.67 |
| 5 | 0.3 | 0.7 | 8.33 | 3.57 | 1.67 | 0.71 |
| 6 | 0.8 | 0.2 | 3.13 | 12.50 | 0.63 | 2.50 |
| 7 | 0.2 | 0.8 | 12.50 | 3.13 | 2.50 | 0.63 |
| 8 | 0.9 | 0.1 | 2.78 | 25.00 | 0.56 | 5.00 |
| 9 | 0.1 | 0.9 | 25.00 | 2.78 | 5.00 | 0.56 |

**Supplementary Figures 1**

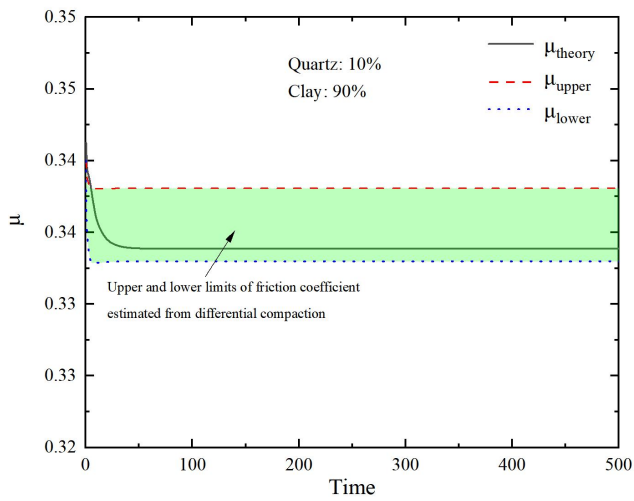

(a)

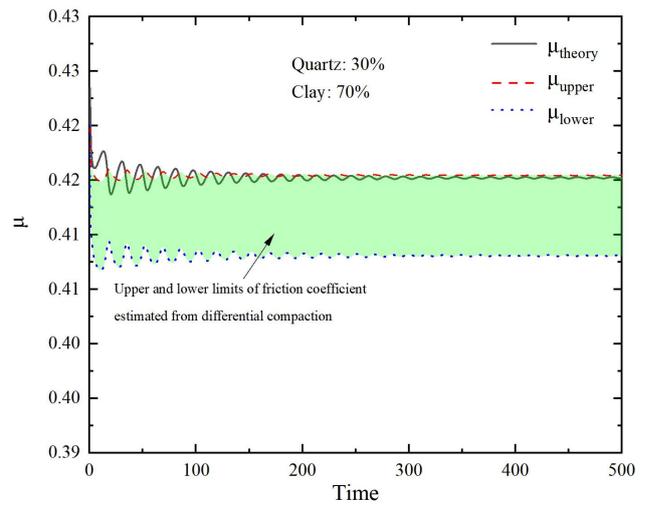

(b)

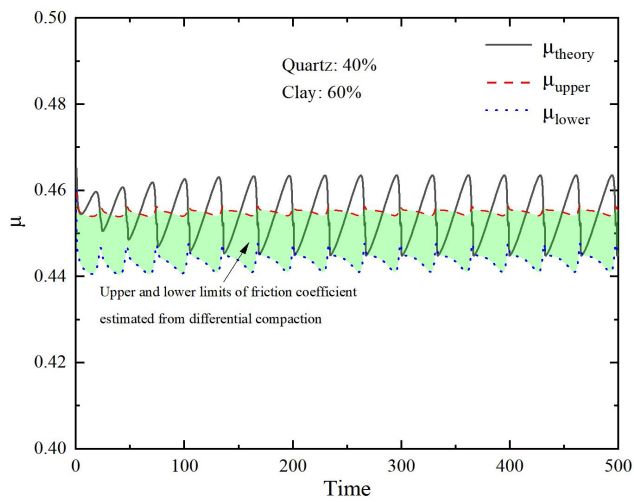

(c)

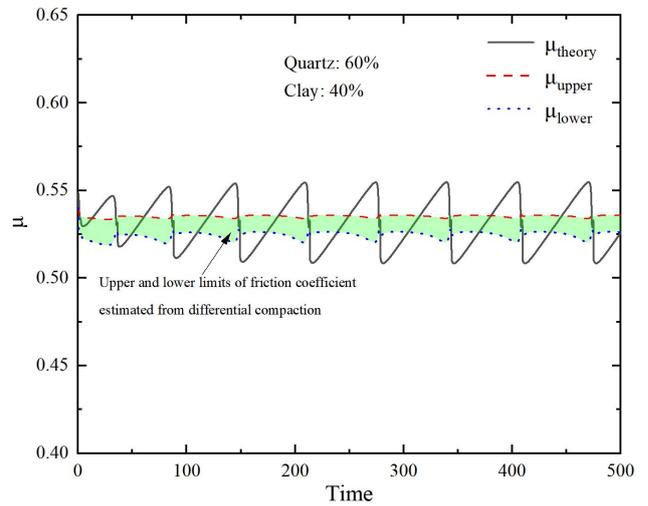

(d)

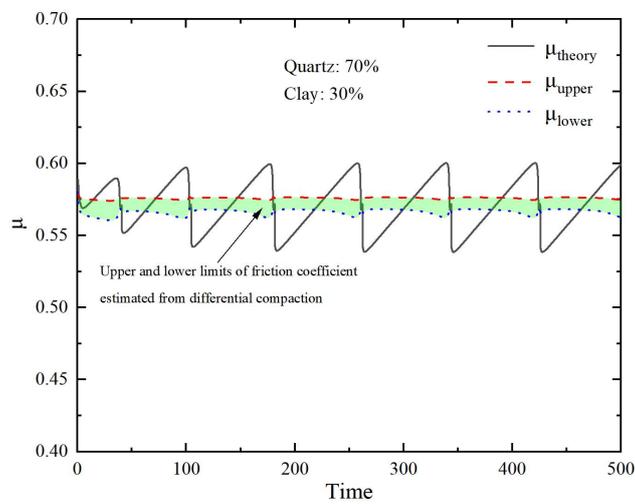

(e)

Supplementary Figure 1: The potential weakening from the differential compaction between the quartz and clay gouges. The constraints with different clay fractions, (a) 90%, (b) 70%, (c) 60%, (d) 40%, and (e) 30%.